\documentclass[11pt,fleqn]{frank}     

\usepackage{amsmath}
\usepackage{amscd, amsthm, amsfonts, amssymb}
\usepackage{mathrsfs, bbm, cancel, xfrac}
\usepackage{marvosym, enumerate, enumitem, upgreek}
\usepackage[all]{xy}
\usepackage[normalem]{ulem}
\usepackage{graphicx}
\usepackage{hyperref, color, microtype}
\usepackage{rotating, lscape, comment, caption}
\usepackage{datetime}
\usepackage[bottom]{footmisc}
\definecolor{darkblue}{rgb}{0,0,1}
\hypersetup{pdfpagemode=UseNone, colorlinks=true, breaklinks=true, linkcolor=darkblue, menucolor=black, urlcolor=darkblue, citecolor=red}

\theoremstyle{remark}

\theoremstyle{definition}
\newtheorem{proposition}{Proposition}

\newtheorem{prob}[proposition]{Problem}

\newtheorem{remark}[proposition]{Remark}
\newtheorem{example}[proposition]{Example}

\def\DHLhksqrt#1#2{%
\setbox0=\hbox{$#1\sqrt{#2\,}$}\dimen0=\ht0
\advance\dimen0-0.2\ht0
\setbox2=\hbox{\vrule height\ht0 depth -\dimen0}%
{\box0\lower0.4pt\box2}}

\allowdisplaybreaks
\renewcommand{\arraystretch}{1.5}

\begin{document}

\sloppy \raggedbottom

\title{The general non-symmetric, unbalanced star circuit\\[0.75ex]
\large On the geometrization of problems in electrical measurement}

\begin{start}

	\author{Christian Eggert}{1}, 
	\author{Ralf G\"aer}{2}, 
	\author{Frank Klinker\thanks{Corresponding author}%
			}{3}\\

	\address{ThyssenKrupp~Rothe~Erde~GmbH, Tremoniastra\ss e 5-11, 44137~Dortmund, Germany\\[0.5ex]
		\href{mailto:christian.eggert@thyssenkrupp.com}{christian.eggert@thyssenkrupp.com}\\}{1}

	\address{Schniewindt~GmbH~\&~Co.~KG, Sch\"ontaler Weg 46, 58809 Neuenrade, Germany\\[0.5ex]
		\href{mailto:ralf.gaeer@schniewindt.de}{ralf.gaeer@schniewindt.de}\\}{2}

	\address{Faculty of Mathematics, TU Dortmund University, 44221 Dortmund, Germany\\[0.5ex]
		\href{mailto:frank.klinker@math.tu-dortmund.de}{frank.klinker@math.tu-dortmund.de}\\}{3}

\noindent
\makebox[0.8\textwidth]{%
\begin{minipage}{0.85\textwidth}
\begin{Abstract}
	We provide the general solution of problems concerning AC star circuits by turning them into geometric problems.
	We show that one problem is strongly related to the Fermat-point of a triangle. We present a solution that is well adapted to the practical application the problem is based on.
	Furthermore, we solve a generalization of the geometric situation and discuss the relation to non-symmetric, unbalanced AC star circuits. 
\end{Abstract}
\end{minipage}}

\renewcommand{\dateseparator}{-}
\let\thefootnote\relax 
\let\thefootnote\relax\footnote{\hspace*{-0.5em}{\em Math.~Semesterber.} {\bf 64} (2017) no.~1, 25-39, \href{https://dx.doi.org/10.1007/s00591-016-0178-8}{doi:10.1007/s00591-016-0178-8}}

\end{start}	
\runningheads{%
{\small\sf Preprint}\hspace*{3.7cm} The general non-symmetric, unbalanced star circuit
}{%
{\small\sf Preprint}\hspace*{7.1cm}   C.~Eggert, R.~G\"aer, F.~Klinker
}



\section{The initial problem: the non-symmetric generator}\label{sec:1}

Nowadays the distribution of power in AC systems is not provided by a single power plant anymore. 
The growth of importance of renewable energies is reflected in an increasing decentralization of energy supply.
To guarantee a stable and continuous operation it is important to constantly and precisely measure the involved currents and voltages.

The question that we discuss in our first part of the text came up during the testing of high voltage generators. 
Its components may independently vary in time, e.g.~due to warming, which results in a non-symmetry of the line voltages.
In the system at hand these line voltages can not be measured directly for technical reasons, but only the phase-to-phase voltages can be measured. 
Therefore, the question was how we can get the first ones from the latter ones.

\begin{figure}[htb]\caption{The basic circuit of a star-shaped generator}\label{abb:schalt-gen}
\centering \includegraphics[scale=0.75]{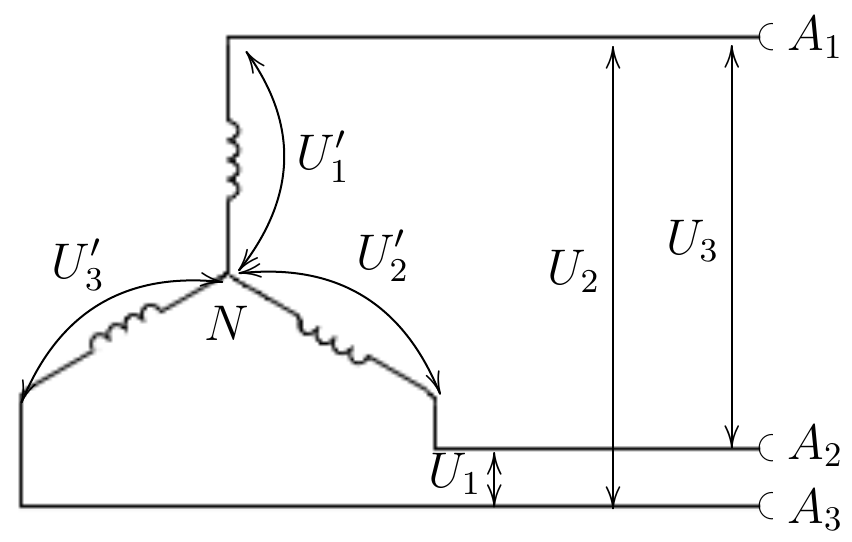}
\end{figure}

A three-phase AC generator typically is star-shaped. 
That means that the three coils of the generator are placed around a turning magnet forming a regular three armed star.
Of each coil one end is grounded ($N$) and the free ends are the phases that form the plug socket ($A_1,A_2,A_3$). 
This situation yields the star circuit as drawn in Figure \ref{abb:schalt-gen}. 

\begin{figure}[htb]\caption{The phasor diagram of the symmetric star-shaped generator}\label{abb:zeiger1}
\centering \includegraphics[scale=0.9]{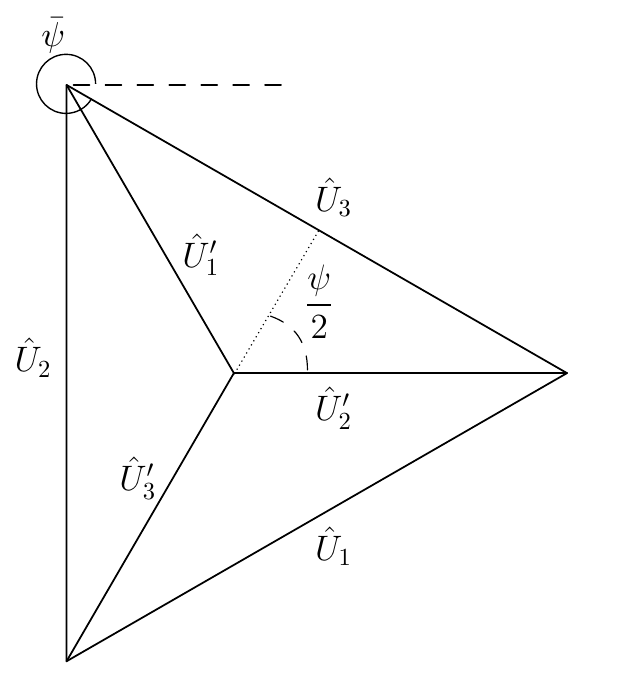}
\end{figure}

The well known symmetric situation is as follows: Between the points $A_i$ and $N$ we have AC voltages with same amplitudes but phase differences $\psi=120^\circ$. Then the voltages can be described in terms of harmonic oscillations in the following way:
\begin{align*}
U_\alpha'&=\hat U_\alpha'\cos(\omega t+\psi_\alpha)=\hat U_\alpha' \Re\left(e^{i\omega t+i\psi_\alpha}\right)\,\quad\alpha=1,2,3\,.
\end{align*}
with $\hat U_\alpha'=\hat U$ for all $\alpha$ and $\psi_1=\psi,\psi_2=0,\psi_3=2\psi$. 
The phase-to-phase voltage $U_3$ between the $A_1$ and $A_2$ is given by the difference of the two voltages $U_1'$ and $U_2'$, i.e.
\begin{align*}
U_3 
&	=\hat U\Re\left(e^{i\omega t+i\psi_2}-e^{i\omega t+i\psi_1}\right)\\	
   &= 2\hat U\sin\frac{\psi_1-\psi_2}{2}\,\cos\Big(\omega t+\frac{\psi_1+\psi_2}{2}-90^\circ\Big)\,.
\end{align*}
Due to the symmetric situation, $|\psi_i-\psi_j|\sim 120^\circ$, $\hat U_i'=\hat U_j'$, the amplitudes of $U_1,U_2$, and $U_3$ are given by
\begin{equation}\label{eq:easy}
\hat U_1= \hat U_2=\hat U_3 =2\hat U\sin60^\circ=\sqrt{3}\,\hat U\,.
\end{equation}
Using the relation between complex numbers and plane geometry, where addition and multiplication are replaced by vector addition and dilatation rotation, we may translate the above circuit into the plane and get the situation from Figure \ref{abb:zeiger1}. 
We emphasize, that in Figure \ref{abb:zeiger1} we only draw the amplitudes of the voltages. 
To see the vector character let $U_1',U_2'$, and $U_3'$ point inwards. Then $U_3=U_1'-U'_2$ points south-east with a phase of $\bar \psi=330^\circ$ as angle between the horizontal and $\hat U_3$ measured in the upper point.\footnote{Whenever we use the term ''voltage'' from now on, we mean the amplitude of the corresponding physical voltage. Therefore, we will omit the $\hat{\ }$ in the notation.}
Such diagrams related to AC calculations are called phasor diagrams and a basic introduction can be found in \cite{SmithAlley,Brookes,Edminister} for example. The main tool for the translation from circuit to phasor diagram is {\em Kirchhoff's mesh rule} or {\em Kirchhoff's voltage law} that states that the sum of the voltages in a closed loop of a circuit vanishes.

The non-symmetric variant of this situation is as follows: The phase differences of the primed line voltages remain $120^\circ$ but their amplitudes differ.
\begin{prob}\label{prob:init}
We start with the star-shaped generator as given in Figure \ref{abb:schalt-gen} with non-symmetric  line voltages (primed). The configuration of our system only allows to measure the phase-to-phase voltages (non-primed). We need a way to compute the primed quantities from the non-primed ones.
\end{prob}

\begin{figure}[htb]\caption{The geometric setup for Problem \ref{prob:geo} of the non-symmetric generator according to Problem \ref{prob:init}}\label{abb:setting}
\centering  \includegraphics[scale=0.7]{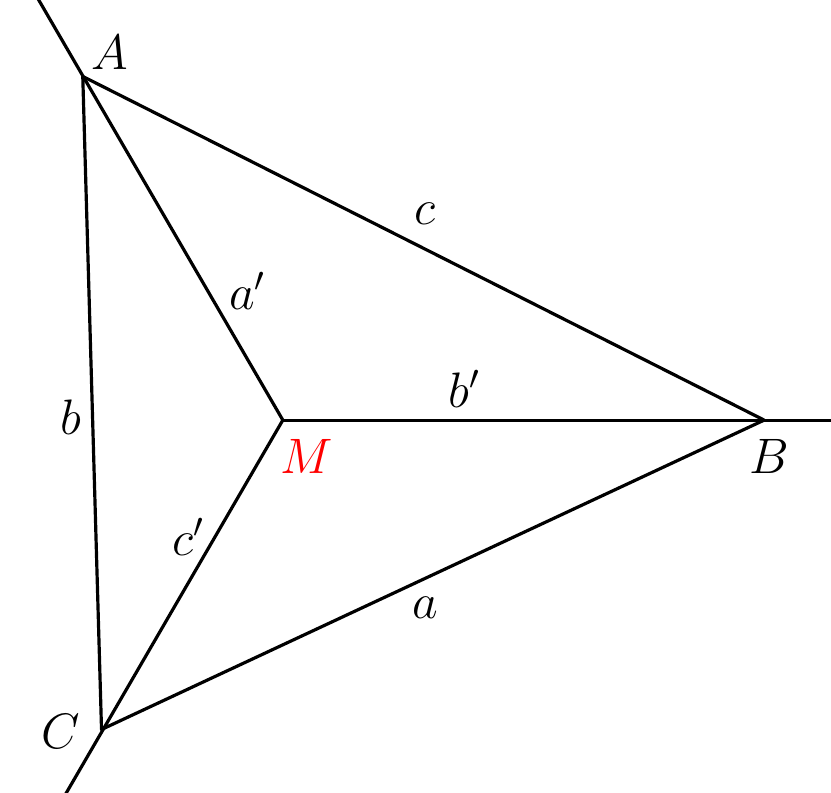}
\end{figure}

The geometric reformulation of Problem \ref{prob:init} is as follows: 

\begin{prob}\label{prob:geo}
Given three rays starting from one point $M$ that pairwise form an angle of $120^\circ$. Furthermore, given three points $A,B,C$ each lying on one ray. These points form a triangle $\triangle(ABC)$, see Figure \ref{abb:setting}.
Starting from this configuration and given the lengths of the three edges $a=|\overline{BC}|$, $b=|\overline{AC}|$,  $c=|\overline{AB}|$ of the triangle, we like to know the lengths $a'$, $b'$, $c'$ of the segments $\overline{MA}$, $\overline{MB}$, $\overline{MC}$.
\end{prob}

\begin{figure}[htb]\caption{Construction of the Fermat-point $M$}\label{abb:construction}
\centering  \includegraphics[scale=0.75]{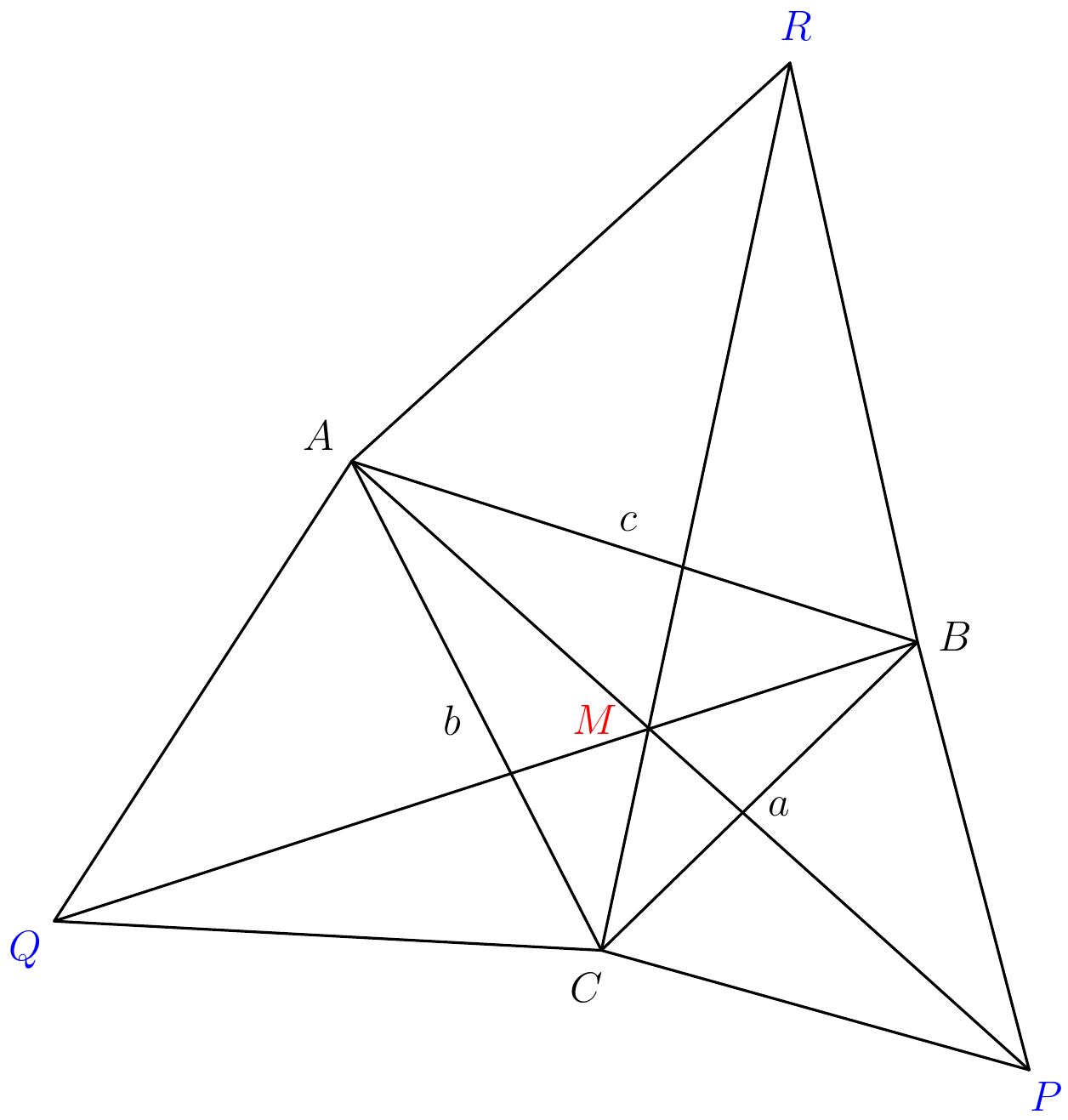}
\end{figure}

The point $M$ that we introduced above is called Fermat-point and gives the solution of a classical geometric problem. The result can be formulated as follows.
\begin{proposition}[The Fermat-point of a triangle]\label{prop:fermat}
Given a triangle $\triangle(ABC)$ with all angles less than $120^\circ$. Then there is a unique point $M$ in the interior for which the lines from $M$ to the corners form equal angles of $120^\circ$. It can be constructed as drawn in Figure \ref{abb:construction} and described as follows:
\begin{enumerate}
\item Over each edge of the triangle $\triangle(ABC)$ draw an equilateral triangle: $\triangle(ARB)$, $\triangle(BPC)$ and $\triangle(ACQ)$.
\item Draw straight lines $\overline{BQ},\overline{AP}$ and $\overline{CR}$.
\item These lines intersect in one point, namely $M$.
\end{enumerate}
\end{proposition}

There exists an elementary geometric proof of Proposition \ref{prop:fermat}. The only things that are used are basic geometric ideas such as congruences of triangles and equality of certain angles. This is the proof presented by Evangelista Torricelli, see \cite{Torri}. It is also published briefly in the English Wikipedia \cite{wiki} and mentioned in \cite{Eriksson}.

The Fermat point in addition has the following very interesting minimizing property. 
\begin{proposition}\label{prop:mini}
For the Fermat point $M$ the sum of the distances to the vertices of the triangle $\triangle(ABC)$ attains its minimum.
\end{proposition} 

This property is not so obvious although there is a very short geometric proof, see \cite{wiki}. 
For a historical survey of the geometric treatment of this problem see \cite{Hofmann} and the wonderful books \cite{AMS,BMS}. In \cite{AMS,BMS} and in \cite{GueronTessler} the authors also explain the mechanical content of the minimizing property that describes the Fermat point as a point of equilibrium, see also Example \ref{exmp:center} for the special situation of an equilateral triangle.

\section{The solution of Problem \ref{prob:geo}: the line voltages of the non-symmetric generator}\label{sec:result}

In this section we give a solution of Problem \ref{prob:init}. This is done by presenting formulas for the quantities $a',b',c'$ from Problem \ref{prob:geo} that are symmetric as functions of $a,b,c$. 

\begin{figure}[htb]\caption{The construction of the Fermat-point: the vector formulation}\label{abb:vectors}\centering
\centering
\includegraphics[width=0.9\textwidth]{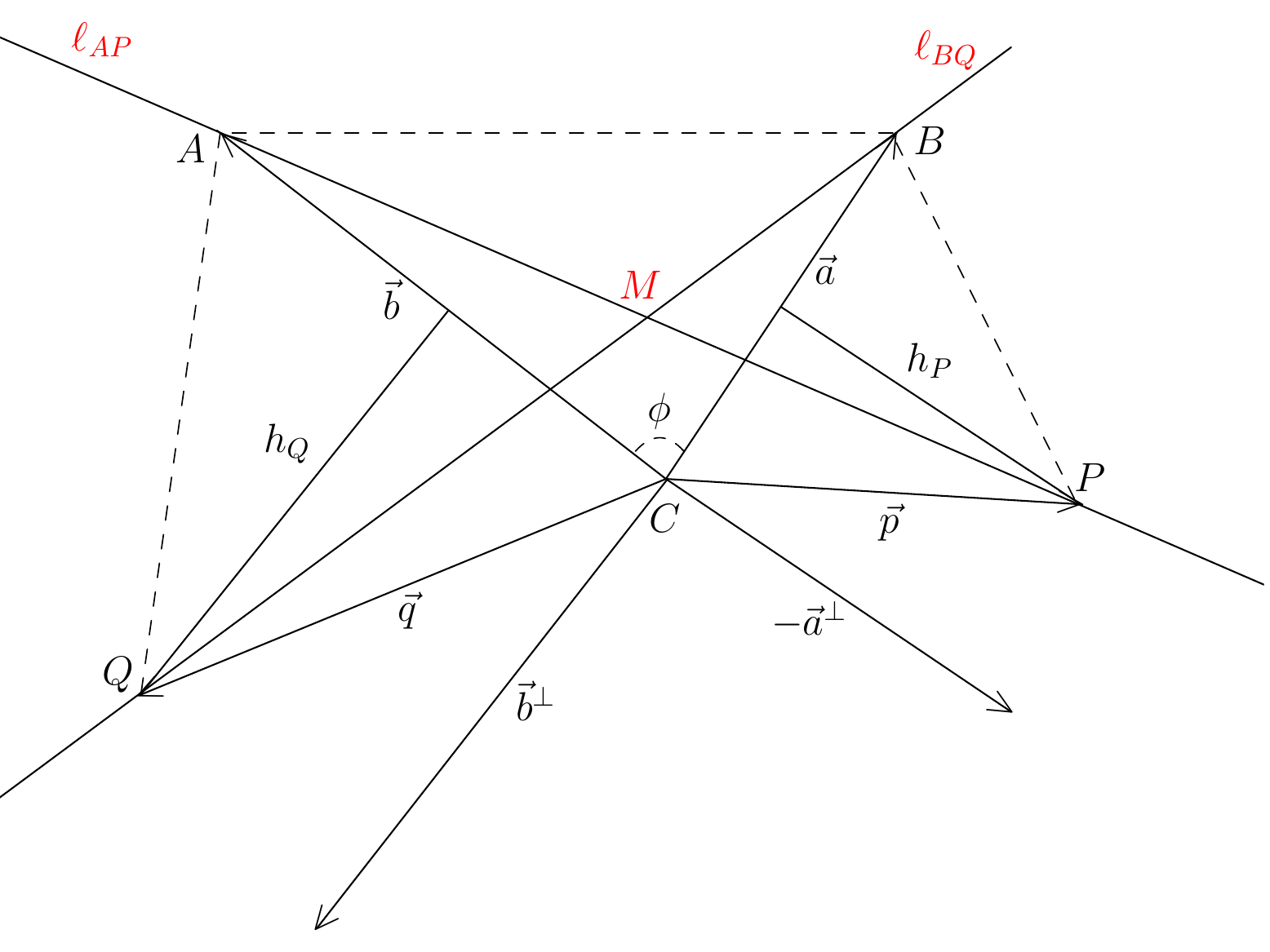}
\end{figure}

As a side result we will get a proof of Proposition \ref{prop:fermat} that explicitly gives the Fermat-point $M$ in terms of the vectors that span the triangle from Figure \ref{abb:setting}. 
In Figure \ref{abb:vectors} we describe this situation by considering $C$ to be the origin of the plane. 

\begin{remark}
When we  take a look at the technical literature we see that explicit calculations are usually performed by using complex numbers. Therefore, the question arises if this would be possible and reasonable here, too. 
Of course, it would be possible. But due to the fact that we look for the intersection of two real lines we would  need to consider real and imaginary parts at some point. 
Geometrically this  means that we would consider all quantities with respect to the standard basis of the euclidean plane. 
In our opinion and concerning to our initial question the use of the vectors that span the triangle is more natural and more reasonable.
In fact, from some point the calculations are almost the same but -- maybe -- a little lengthier when we would use complex numbers.
\end{remark}

Before starting the calculations we recall some useful facts about vectors of the euclidean plane.\footnote{For more details on basics in linear algebra see \cite{Lorenz,Lorenz2}, for example.}
We consider two non vanishing plane vectors ${\vec a}$ and  ${\vec b}$ drawing an angle $\phi=\measuredangle(\vec a,\vec b)$.\footnote{
By $\measuredangle(\vec a,\vec b)$ we will always mean the oriented angle $0^\circ\leq\phi<360^\circ$ that goes from 
$\vec a$ in counterclockwise direction to ${\vec b}$. The angle between $\vec b$ and $\vec a$ is then $\measuredangle(\vec b,\vec a)=360^\circ-\phi$ if $\phi\neq 0^\circ$ and $\measuredangle(\vec b,\vec a)=0^\circ$ if $\phi=0^\circ$. If we do not care about the orientation we write e.g.\ $\measuredangle(\vec a,\vec b)\sim 45^\circ$, i.e. $\measuredangle(\vec a,\vec b)\sim45^\circ\sim 315^\circ$.}
It is given by $\langle{\vec a},{\vec b}\rangle=ab \cos\phi$ with $\langle \cdot,\cdot  \rangle$ being the euclidean product and $a:=\|{\vec a}\|=\sqrt{\langle {\vec a},{\vec a}\rangle}$.

For any vector ${\vec a}$ there exists an unique perpendicular vector ${\vec a}^\perp$ that has length as $\|\vec{a}^\perp\|=a$ and both vectors form a positive basis of the plane w.r.t.\ the order $\{\vec{a},\vec{a}^\perp\}$. 
We have $\langle {\vec a}^\perp,{\vec b} \rangle= -\langle {\vec a},{\vec b}^\perp\rangle=  a b \sin\phi$.

In our situation ${\vec a}$ and ${\vec b}$ are linearly independent such that we can expand ${\vec a}^\perp$ and ${\vec b}^\perp$ as linear combinations. With ${\vec a}^\perp=\alpha {\vec a}+\beta {\vec b}$ we get $\langle {\vec a}^\perp,{\vec b}^\perp\rangle  =\alpha \langle {\vec a},{\vec b}^\perp\rangle$ and $\langle {\vec a}^\perp,{\vec a}^\perp\rangle  =\beta \langle {\vec b},{\vec a}^\perp\rangle$. This yields  
$\alpha = -\cot\phi$ and $\beta = \frac{a}{b\sin\phi}
$. 
Doing the same for ${\vec b}^\perp$ we get 
\begin{align}
{\vec a}^\perp &= \frac{a}{\sin\phi}\big(
	-\cos\phi \frac{\vec a}{a} + \frac{\vec b}{b}\big)
\,, \quad
{\vec b}^\perp = \frac{b}{\sin\phi}\big( 
	\cos\phi\,\frac{\vec b}{b} - \frac{\vec a}{a} \big)
\,.
\end{align}

Let us turn to our situation from Figure \ref{abb:vectors} and include the additional perpendicular vectors $\vec a^\perp$ and ${\vec b}^\perp$ into our discussion. 
We note that the lengths $h_P$ and $h_Q$ of the two heights of the equilateral triangles are given by $h_P=\frac{\sqrt{3}}{2}a$ and $h_Q=\frac{\sqrt{3}}{2}b$. With $\sin 60^\circ=\frac{\sqrt{3}}{2}$, $\cos60^\circ=\frac{1}{2}$  and $\sqrt{3}\cos\phi +\sin \phi= 2\sin(\phi+60^\circ)$ the position vectors $\vec p$ and $\vec q$ of $P$ and $Q$ are given by
\begin{equation}
\begin{aligned}
\vec{p} &=\frac{1}{2}{\vec a} -\frac{\sqrt{3}}{2} {\vec a}^\perp 
= \frac{\sin(\phi+60^\circ)}{\sin\phi} {\vec a}  
				-\frac{\sqrt{3}a}{2b\sin\phi} {\vec b}
\,,\\	
\vec{q} &=\frac{1}{2}{\vec b} +\frac{\sqrt{3}}{2} {\vec b}^\perp 
	= \frac{\sin(\phi+60^\circ)}{\sin\phi}{\vec b} 
				- \frac{\sqrt{3}b}{2a\sin\phi} {\vec a}\,. 
\end{aligned}
\end{equation}
Therefore, the lines $\ell_{AP}$ and $\ell_{BQ}$ that contain the segments $\overline{AP}$ and $\overline{BQ}$ are parametrized by
\begin{equation}
\begin{aligned}
\ell_{AP}(\tau)	&= {\vec b}+\tau (\vec{p}-{\vec b}) 
			=\Big(1-\tau-\frac{\sqrt{3}a}{2b\sin\phi}\tau \Big){\vec b}
			  +\frac{\sin(\phi+60^\circ)}{\sin\phi}\tau\, {\vec a} 
			  \,,\\
\ell_{BQ}(\sigma)&= {\vec a}+\sigma (\vec{q}-{\vec a}) 
			= \Big(1-\sigma-\frac{\sqrt{3}b}{2a\sin\phi}\sigma\Big){\vec a} 
			   +\frac{\sin(\phi+60^\circ)}{\sin\phi}\sigma\,{\vec b} 
			   \,.
\end{aligned}
\end{equation}
The intersection point of $\ell_{AP}$ and $\ell_{BQ}$ is determined by the solution $(\tau_0,\sigma_0)$ of the equation $\ell_{AP}(\tau)=\ell_{BQ}(\sigma)$ that is equivalent to 
\begin{align*}
& \left(\frac{\sin(\phi+60^\circ)}{\sin\phi}\tau
				-\Big(1-\sigma-\frac{\sqrt{3}b}{2a\sin\phi}\sigma\Big)\right){\vec a} \\
	 & \qquad  = \left( \frac{\sin(\phi+60^\circ)}{\sin\phi}\sigma
			-\Big(1-\tau-\frac{\sqrt{3}a}{2b\sin\phi}\tau \Big)\right){\vec b} \\
\iff& \underbrace{\begin{pmatrix}
	\sin(\phi+60^\circ)
		& \sin\phi+\frac{\sqrt{3}b}{2a}\\
	\sin\phi +\frac{\sqrt{3}a}{2b}	
		& \sin(\phi+60^\circ)
\end{pmatrix}}_{\displaystyle =\Omega} \begin{pmatrix}\tau\\\sigma\end{pmatrix}
	= \sin\phi \begin{pmatrix}1\\1\end{pmatrix}\,.
\end{align*}
The determinant of the coefficient matrix is 
\begin{align*}
\det\Omega
=\ & 
= -\frac{\sqrt{3}\sin\phi}{2ab}\left( a^2+b^2-2ab\cos(\phi+60^\circ) \right)
\end{align*}  
such that the solution $(\tau_0,\sigma_0)$ is given by 
\begin{align}
\begin{pmatrix}\tau_0\\\sigma_0\end{pmatrix}
&=\frac{-\sin\phi}{\det\Omega}
\begin{pmatrix}
	-\sin(\phi+60^\circ) 		& \sin\phi+\frac{\sqrt{3}b}{2a}\\
	\sin\phi +\frac{\sqrt{3}a}{2b}	& -\sin(\phi+60^\circ)
\end{pmatrix}
\begin{pmatrix}1\\1\end{pmatrix}\nonumber\\
&=\begin{pmatrix}
\displaystyle \frac{\sqrt{3}\, b^2+2ab\sin(\phi-60^\circ)}%
					{\sqrt{3}\big(a^2+b^2\big)-2\sqrt{3}\,ab\cos(\phi+60^\circ)}\\[2ex]
\displaystyle \frac{\sqrt{3}\, a^2+2ab\sin(\phi-60^\circ)}%
					{\sqrt{3}\big(a^2+b^2\big)-2\sqrt{3}\,ab\cos(\phi+60^\circ)}
\end{pmatrix}
\end{align} 

We use this result to calculate the position vector $\vec m$ of the Fermat-point $M$ in the situation of Figure \ref{abb:vectors}.
\begin{equation}
\vec m= \ell_{AP}(\tau_0)=\ell_{BQ}(\sigma_0)
= (1-\tau_0)\vec{b}+\tau_0\vec p = (1-\sigma_0)\vec{a}+\sigma_0\vec q \,.
\end{equation}
We use
$\langle \vec b,\vec p\rangle 
= \frac{1}{2}\langle \vec b,\vec a\rangle-\frac{\sqrt{3}}{2}\langle \vec b,\vec a^\perp\rangle
= ab\cos(\phi+ 60^\circ)$ 
such that in terms of $\tau_0$
\begin{align}
&\begin{aligned}
(c')^2 & = \|\vec m\|^2 
 =  (1-\tau_0)^2b^2+\tau_0^2 \|\vec p\|^2+2\tau_0(1-\tau_0)\langle \vec b,\vec p\rangle
\\
& = (1-\tau_0)^2b^2+\tau_0^2 a^2+2\tau_0(1-\tau_0) ab\cos(\phi+ 60^\circ)\,.\label{c'}
\end{aligned}
\end{align}
Similarly we get 
\begin{align}
(a')^2 & =|AM|^2
= \tau_0^2\big(a^2+b^2-2ab\cos(\phi+60^\circ)\big)\label{a'}\,,\\
(b')^2 & = |BM|^2
= \sigma_0^2\big(a^2+b^2-2ab\cos(\phi+60^\circ)\big)\label{b'}\,.
\end{align}
To get formulas for $a',b'$, and $c'$ that are symmetric with respect to $a,b$, and $c$ we recall the cosine-theorem that says 
\begin{equation}\label{eq:costh}
2ab\cos\phi=a^2+b^2-c^2\,.
\end{equation}
In the same way we express $2ab\sin\phi$ as
\begin{equation}\label{eq:sinth}
4a^2b^2\sin^2\phi = 4a^2b^2(1-\cos^2\phi)
=   4a^2b^2-(a^2+b^2-c^2)^2
= \Uptheta^4
\end{equation}
where we use the abbreviation
\begin{equation}\label{eq:delta}
\Uptheta^2=\sqrt{(c+a+b)(a+c-b)(b+c-a)(a+b-c)}
\end{equation}
that is invariant under relabeling the three edges.\footnote{
Formulas \eqref{eq:sinth} with \eqref{eq:delta} recall the famous Heron formula that gives the area of an triangle in terms of its three edges. This area is given by 
$4\cdot \text{area}\big(\triangle(ABC)\big)=\Uptheta^2$.}

The denominator of $a',b'$, and $c'$ agrees -- up to a factor of 3 -- with the denominator of $\tau_0$ and $\sigma_0$. It can be written as
\begin{align*}
a^2+  b^2-2ab\cos(\phi+60^\circ)
=\ & \frac{1}{2}(a^2+b^2+c^2) +\frac{\sqrt{3}}{2}\Uptheta^2\,.
\end{align*}
The numerator of $a'$ is given by the numerator of $\tau_0^2$. 
It differs from that of $b'$ or $\sigma_0^2$ only by interchanging $a$ and $b$ and is given by 
\begin{align*}
\left(\sqrt{3}  b^2  +2ab\sin(\phi-60^\circ)\right)^2 
&= \frac{1}{4}\left(\sqrt{3}(b^2+c^2-a^2)+\Uptheta^2\right)^2 \,.
\end{align*}
We insert these expressions into \eqref{a'} and \eqref{b'} and  get expressions for $a'$ and $b'$. 
A similar calculation yields the remaining length $c'$ from \eqref{c'}. 
\begin{proposition}\label{prop:result}
Given a triangle $\triangle(ABC)$ and the Fermat-point $M$ as given in Figure \ref{abb:setting}. Then $a',b'$, and $c'$ are given in terms of $a,b$, and $c$ by 
\begin{align}
(a')^2 &
= \frac{1}{6}\cdot\frac{\left(\sqrt{3}(b^2+c^2-a^2)+\Uptheta^2\right)^2 }{a^2+b^2+c^2 +\sqrt{3}\,\Uptheta^2}\,,\label{eqa'}\\
(b')^2 &
= \frac{1}{6}\cdot\frac{\left(\sqrt{3}(a^2+c^2-b^2)+\Uptheta^2\right)^2 }{a^2+b^2+c^2 +\sqrt{3}\,\Uptheta^2}\,,\label{eqb'}\\
(c')^2 &
= \frac{1}{6}\cdot\frac{\left(\sqrt{3}(a^2+b^2-c^2)+\Uptheta^2\right)^2 }{a^2+b^2+c^2 +\sqrt{3}\,\Uptheta^2}\,.\label{eqc'}
\end{align}
\end{proposition}

This also yields the solution of initial Problem \ref{prob:init} but we postpone the formulation to the summary, see Section \ref{sec:concl}.

\begin{example}\label{exmp:center}
As a first example and also a first check of our result we consider $a=b=c$, i.e.\ an equilateral triangle. In this case we have $\Uptheta^2=\sqrt{3}a^2$ and $a'=b'=c'=\frac{1}{\sqrt{3}}a$ which is exactly the result from \eqref{eq:easy}.  In particular $\vec m=\frac{1}{3}(\vec a+\vec b)$ is the position vector of the geometric center of the triangle.
\end{example}

\begin{remark}
To finish a proof of Proposition \ref{prop:fermat} we have to do some more calculations:

Firstly, we have to show that $M$ lies on the line that connects the origin and $R$, i.e. there exists $\lambda_0$ such that  $\vec m=\ell_{CR}(\lambda_0)=\lambda_0\vec r=\frac{\lambda_0}{2}\left(\vec a+\vec b+\sqrt{3}(\vec a^\perp-\vec b^\perp)\right)$. 

Secondly, we have to show that the angles $\measuredangle(\vec p-\vec a,\vec q-\vec b)$,  $\measuredangle(\vec m,\vec p-\vec a)$, and $\measuredangle(\vec m,\vec q-\vec b)$ coincide and, therefore, are given by $120^\circ$. 
\end{remark}

\begin{remark}
Starting from our results \eqref{eqa'}-\eqref{eqc'} we can prove the minimizing property from Proposition \ref{prop:mini}. 
For this we look for critical points of the function $f(a',b',c',\psi,\phi)=a'+b'+c'$  w.r.t.\ the three constraints 
$g_x(x,y,z,\phi,\psi)=y^2+z^2-2yz\cos\phi -a^2=0$, 
$g_y(x,y,z,\phi,\psi)=x^2+z^2-2xz\cos\psi -b^2=0$, and 
$g_z(x,y,z,\phi,\psi)=x^2+y^2-2xy\cos(\phi+\psi)-c^2=0$ by using the Lagrange method. 

We have to show that our solution yields a critical point of the Lagrange function $L=f+\sum_{\omega\in\{x,y,z\}}\lambda_\omega g_\omega$ for Lagrange parameters  $2(xy+xz+yz)\lambda_\omega=\omega$.

Moreover, we have to show that this critical point is indeed a minimum, for example by using the rendered Hessian, see \cite{HassellRees}.
\end{remark}

\section{A generalization: the unbalanced star circuit}

\begin{figure}[htb]\caption{The unbalanced star circuit}\label{fig:nonbalanced}
\centering\includegraphics[scale=1]{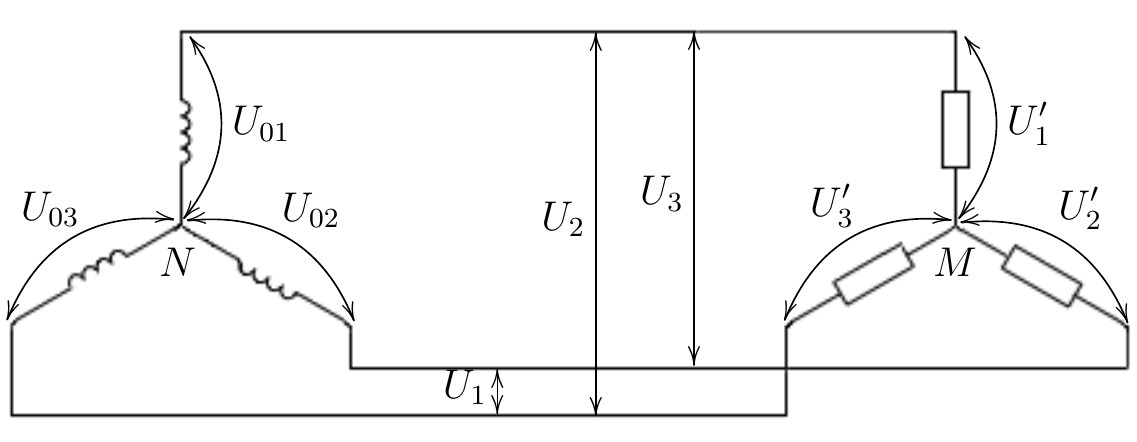}
\end{figure}

We consider a 3-phase AC star circuit with unbalanced star point, i.e.\ the line between the star point $N$ of the generator and the star point $M$ of the circuit is missing, see Figure \ref{fig:nonbalanced}. 

Let us first assume that the perfect generator provides three equal line-voltages $U_{01}=U_{02}=U_{03}=\frac{1}{\sqrt{3}}U$ with a phase difference of $120^\circ$. Then the phase-to-phase voltages fulfill $U_1=U_2=U_3=U$ and they have a phase difference of $120^\circ$, too, resulting in an equilateral phasor diagram. 

\begin{figure}[htb]\caption{The phasor diagram of the load of an unbalanced star circuit for a symmetric generator}\label{fig:phasor-2}
\centering\includegraphics[scale=0.8]{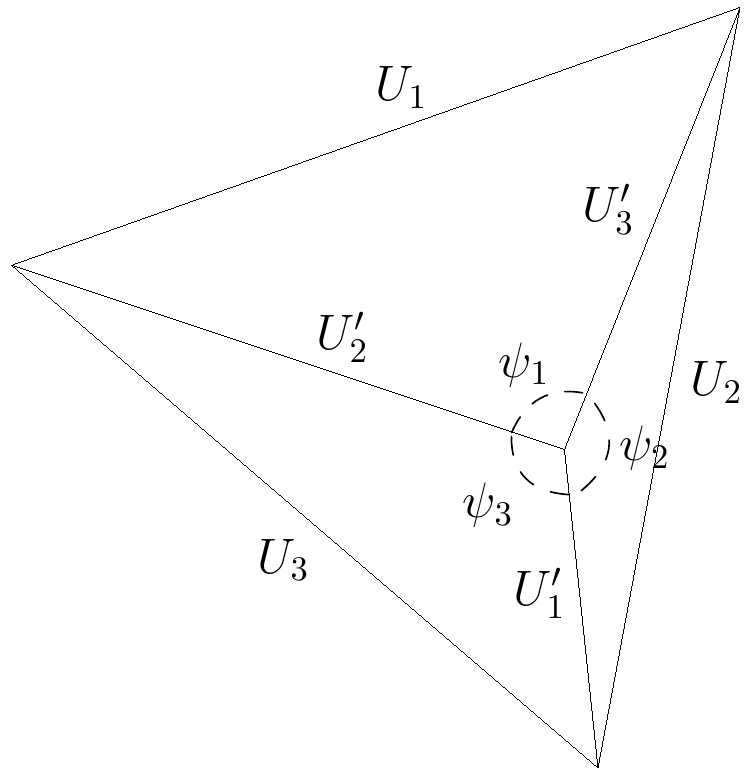}
\end{figure}

The unbalanced configuration typically yields $U_1'\neq  U_2'\neq  U'_3$ for the primed line voltages of the load. This is reflected in the phasor diagram in such a way that the star point is displaced in the equilateral triangle defined by $U_1,U_2,U_3$, see Figure \ref{fig:phasor-2}. We emphasize the fact that the phase-to-phase voltages of the generator and the load are the same due to the mesh rule.
Such unbalanced star circuits have been considered in  \cite{Rathore} for special almost symmetric configurations, e.g.\ $ U_1'= U_2'$.

We will now consider non-equal phase-to-phase-voltages $U_\alpha$ that are provided by a non-symmetric generator according to Section \ref{sec:1}.
We now ask the following question: 
Knowing the phase-to-phase voltages of the load/generator we want to recover the primed line voltages of the load.

Of course, the phase-to-phase voltages alone do not contain enough information to obtain a solution. Typically, you know about the technical configuration of the load, for example about resistors, capacities or inductances, see \cite{Rathore}. 
To create a purely geometric problem, we assume the load to be a black box of which we do not know the exact components but we know about the phase differences of the primed voltages.
\begin{prob}\label{prob:unbalanced}
Given the phase-to-phase-voltages $U_1,U_2,U_3$ of the load provided by a non-symmetric generator as well as the phase-differences $\psi_1$ and $\psi_2$ of the primed line voltages of the load: What are the values of the line voltages $U_1'$, $U_2'$, and $U_3'$?
\end{prob}
The geometric reformulation is as follows. 
\begin{prob}\label{prob:general}
Given a plane triangle $\triangle(ABC)$ with lengths $a,b,c$ of its edges. 
Furthermore, given an unknown point $X$ in the interior of $\triangle(ABC)$ of which we know the angles $\psi_a=\angle(BXC)$ and $\psi_b=\angle(CXA)$: What are the lengths of the connecting edges $a'=|AX|,b'=|BX|$, and $c'=|CX|$?
\end{prob}

In fact, for phase differences $\psi_a=\psi_b=\psi_c= 120^\circ$ this yields another formulation of Problem \ref{prob:init} in terms of load instead of generator.

\begin{figure}[htb]\caption{The geometric description of the load of an unbalanced star circuit with non-symmetric generator}\label{fig:vector2}
\centering
\includegraphics[width=\textwidth]{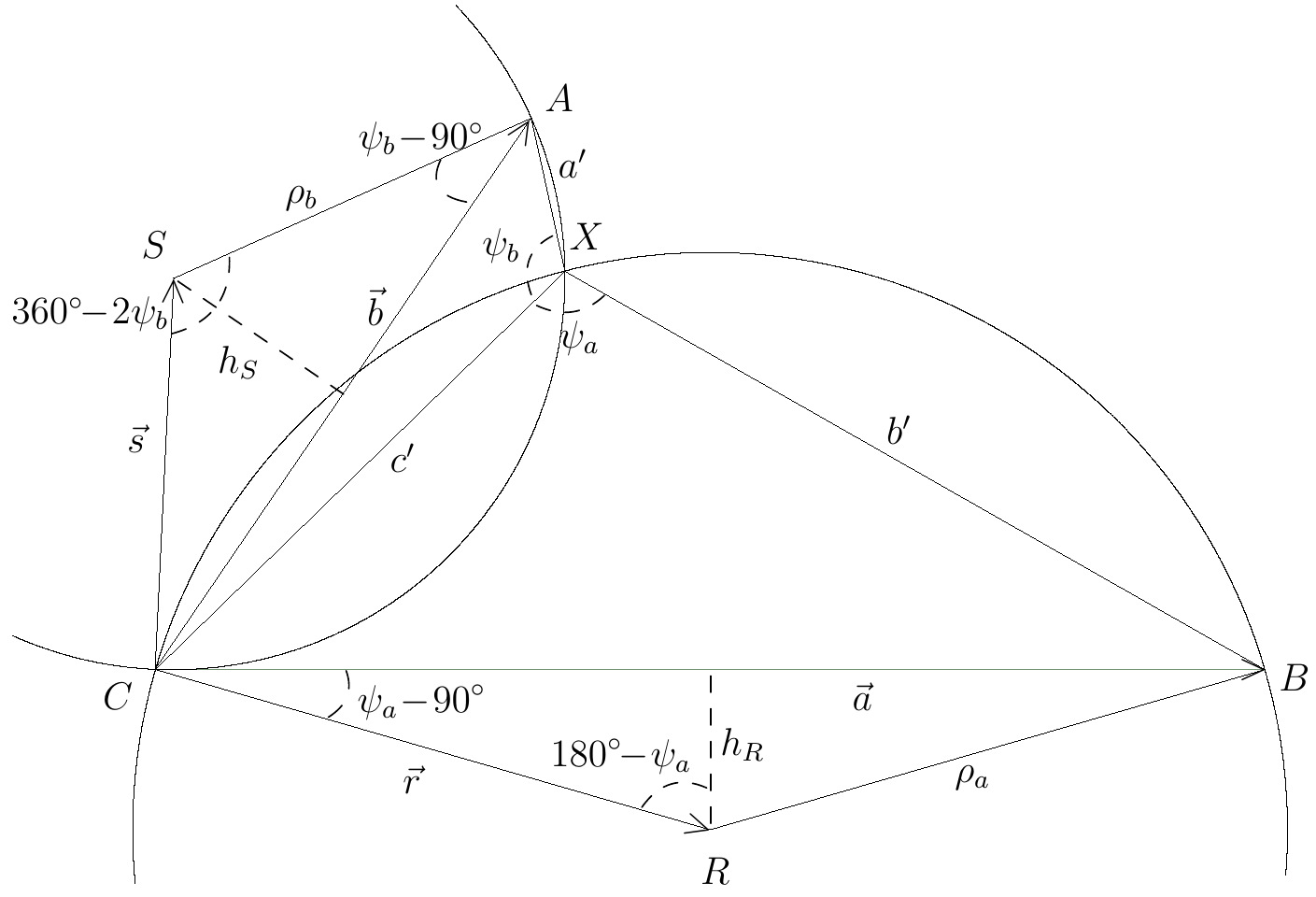}
\end{figure}

For the discussion of Problem \ref{prob:general} we consider Figure \ref{fig:phasor-2} with non equal phase-to-phase voltages and translate it to the vector picture from Figure \ref{fig:vector2}. 
We will make use of the preliminaries and the notation from Section \ref{sec:result} and add a few more quantities that we describe next.

Due to the inscribed angle theorem, all points $X$ that draw an angle $\psi_b$ with the endpoints of the segment $\overline{AC}$ lie on a circle with center $S$, the circumcircle of $\triangle(XAC)$, see \cite{AgricolaFriedrich} for example. Suppose $\psi_b\geq90^\circ$ then $S$ and $X$ lie on different sides of $\overline{AC}$ and the central angle is given by $2(180^\circ-\psi_b)=360^\circ-2\psi_b$. 
If the angle $\psi_a$ obeys the restriction $\psi_a\geq 90^\circ$, too, the point $X$ is the intersection of the two circles with centers $R$ and $S$ that contain the two chords $\overline{AC}$ and $\overline{BC}$, respectively.\footnote{The restriction on the two angles $\psi_a,\psi_b$ before is actually no restriction, because due to $\psi_a,\psi_b,\psi_c<180^\circ$ at least two of the three angles $\psi_a,\psi_b$, and $\psi_c=360^\circ-\psi_a-\psi_b$ are of this form. Therefore, Figure \ref{fig:vector2} describes the general situation, at least after renaming the points and edges of the triangle.}

The radii of the circumcircles of $\triangle(XCB)$ and $\triangle(XAC)$ are given by  
$\rho_a=\frac{a}{2\cos(\psi_a-90^\circ)}=\frac{a}{2}\textstyle\sqrt{1+\cot^2\psi_a}$ and  $\rho_b=\frac{b}{2}\textstyle\sqrt{1+\cot^2\psi_b}$, respectively. 
The heights of the corresponding triangles $\triangle(BCR)$ and $\triangle(ASC)$ are  
$ h_R=\frac{a}{2}\tan(\psi_a-90^\circ)=-\frac{a}{2}\cot\psi_a$ and 
$ h_S=-\frac{b}{2}\cot\psi_b$, respectively. Therefore, the  position vectors of the centers of the circumcircles are 
$\vec r=\frac{1}{2}\vec a-\frac{h_R}{a}\vec a^\perp=\frac{1}{2}(\vec a+\cot\psi_a\,\vec a^\perp)$ and  
$\vec s=\frac{1}{2}(\vec b-\cot\psi_b\,\vec b^\perp)$.

We will calculate the position vector $\vec x$ of $X$ whose length is given by $c'$. For this, we write $\vec x$ as a linear combination of the two vectors that span the triangle:
\[
\vec x=\frac{\alpha}{2}\,\vec a+\frac{\beta}{2}\,\vec b\,.
\]
As said before, $X$ is given as an intersection point of the two circles
\[
\big\{\vec y\ \big|\,\|\vec y-\vec r\|^2=\rho_a^2\big\}\ \text{ and  }\ 
\big\{\vec y\ \big|\,\|\vec y-\vec s\|^2=\rho_b^2\big\} 
\]
such that the coefficients of $\vec x$ obey
\begin{align*}
&\left\{
\begin{aligned}
&\big\| (\alpha-1)\vec a -\cot\psi_a\,\vec a^\perp +\beta\,\vec b\big\|^2
		=a^2(1+\cot^2\psi_a)\\
&\big\| (\beta-1)\vec b +\cot\psi_b\,\vec b^\perp +\alpha\,\vec a\big\|^2
		=b^2(1+\cot^2\psi_b)\\
\end{aligned}
\right.
\\
\Leftrightarrow
&\left\{\begin{aligned}
0&=\alpha^2a^2+\beta^2b^2+2\alpha\beta\langle\vec a,\vec b\rangle -2\alpha a^2
-2\beta\langle\vec a,\vec b\rangle-2\beta \cot\psi_a\langle \vec a^\perp,\vec b\rangle
\\
0&=\alpha^2a^2+\beta^2b^2+2\alpha\beta\langle\vec a,\vec b\rangle -2\beta b^2
-2\alpha\langle\vec a,\vec b\rangle-2\alpha \cot\psi_b\langle \vec a^\perp,\vec b\rangle
\end{aligned}
\right.
\end{align*}
We subtract the two equations and get 
\begin{align*}
&\alpha\Big( 
a^2-\langle\vec a,\vec b\rangle-\cot\psi_b\langle\vec a^\perp,\vec b\rangle
\Big)
-\beta\Big(
b^2-\langle\vec a,\vec b\rangle-\cot\psi_a\langle\vec a^\perp,\vec b\rangle
\Big) = 0\,.
\end{align*}
We write this as $\beta=t\alpha$ with  
$t=\frac{a^2-\langle\vec a,\vec b\rangle-\cot\psi_b\langle\vec a^\perp,\vec b\rangle}{b^2-\langle\vec a,\vec b\rangle-\cot\psi_a\langle\vec a^\perp,\vec b\rangle}$. 
We introduce the length of the third edge of the triangle, $c=\|\vec a-\vec b\|$, and use
$2\langle\vec a,\vec b\rangle =a^2+b^2-c^2$
as well as
$2\langle \vec a^\perp,\vec b\rangle = \Uptheta^2$, see \eqref{eq:costh}-\eqref{eq:delta}, and write
\begin{equation}
\begin{aligned}
t&=\frac{c^2+a^2-b^2 -\cot\psi_b\Uptheta^2}{c^2+b^2-a^2-\cot\psi_a\Uptheta^2}\,,&
t_*&=\frac{c^2+b^2-a^2-\cot\psi_a\Uptheta^2}{c^2+a^2-b^2 -\cot\psi_b\Uptheta^2}\,.
\end{aligned}
\end{equation}
We insert this into the quadratic equations and get for $\alpha,\beta\neq 0$ 
\begin{equation}
\begin{aligned}
\alpha&=\frac{2a^2+t(a^2+b^2-c^2)+t\cot(\psi_a)\Uptheta^2}{a^2+t^2 b^2+t(a^2+b^2-c^2)}\,,\\
\beta&=\frac{2b^2+t_*(a^2+b^2-c^2)+t_*\cot(\psi_b)\Uptheta^2}{b^2+t_*^2 a^2+t_*(a^2+b^2-c^2)}\,.\\
\end{aligned}
\end{equation}
The length of $\vec x$, 
\[
(c')^2
=\frac{1}{4}\left( \alpha^2 a^2+\beta^2b^2+2\alpha\beta\langle\vec a,\vec b\rangle\right)
=\frac{1}{4}\left( \alpha^2 a^2+\beta^2b^2+\alpha\beta(a^2+b^2-c^2)\right)\,,
\]
is now obtained by a lengthy calculation. In particular, we use
\[
1-\cot\psi_a\cot\psi_b=-\cot(\psi_a+\psi_b)(\cot\psi_a+\cot\psi_b)=\cot\psi_c(\cot\psi_a+\cot\psi_b)\,.
\] 
The result is formulated in the next Proposition.

\begin{proposition}

We consider the situation from Problem \ref{prob:general}. Then the length  $c'$ of the connecting edge is given  by

\hspace*{-2.3em}\begin{minipage}{1.07\textwidth}
\begin{align}\begin{aligned}
(c')^2
&=\frac{\frac{1}{4}\left((1-\cot\psi_a\cot\psi_b)\Uptheta^2
-(\cot\psi_a+\cot\psi_b)(a^2+b^2-c^2)\right)^2}{\big(c^2+a^2\cot^2\psi_a+b^2\cot^2\psi_b\big)
+\cot\psi_a\cot\psi_b(a^2+b^2-c^2)-(\cot\psi_a+\cot\psi_b)\Uptheta^2} 
\\
&=\frac{\frac{1}{4}\big(\cot\psi_a+\cot\psi_b\big)^2\big((a^2+b^2-c^2)-\Uptheta^2\cot\psi_c\big)^2}%
{a^2(1+\cot^2\psi_a) +b^2(1+\cot^2\psi_b)
-(\cot\psi_a+\cot\psi_b)\big((a^2+b^2-c^2)\cot\psi_c+\Uptheta^2\big)}
\,.
\end{aligned}\label{eq:c-gen}
\end{align}
\end{minipage}
\end{proposition}
By interchanging the roles of $a,b$, and $c$ we get the results for $a'$ and $b'$. To end up this section we will check our result by discussing some special examples:
\begin{itemize}
\item The equilateral triangle with $a=b=c$ yields $\Uptheta^2=\sqrt{3}a^2$ and
\[
\begin{aligned}
c'&=\frac{a}{2}\cdot\frac{(1-\cot\psi_a\cot\psi_b)\sqrt{3}
-(\cot\psi_a+\cot\psi_b)}{\sqrt{1+\cot^2\psi_a+\cot^2\psi_b +\cot\psi_a\cot\psi_b -\sqrt{3}(\cot\psi_a+\cot\psi_b)}} \\
&=\frac{a}{2}\cdot\frac{(\cot\psi_a+\cot\psi_b)(\sqrt{3}\cot\psi_c-1)}%
{\sqrt{2+\cot^2\psi_a+\cot^2\psi_b
-(\cot\psi_a+\cot\psi_b)\big(\cot\psi_c+\sqrt{3}\big)}}
\,.
\end{aligned}
\]
\item The situation of equal angles, $\psi_a=\psi_b=\psi_c=120^\circ$, with  $\cot\psi_a=-\frac{\sqrt{3}}{3}$ yields
\[
(c')^2
=\frac{\frac{1}{6}\left(\Uptheta^2
+\sqrt{3}(a^2+b^2-c^2)\right)^2}{c^2+a^2+b^2+\sqrt{3}\Uptheta^2} 
\]
which is exactly the result we obtained in \eqref{eqc'}.
\item 
The isosceles triangle with $a=b$, $\psi_a=\psi_b$ yields $\Uptheta^2=c\sqrt{4a^2-c^2}$. 
This is the case mainly discussed in \cite{Rathore}. 
In this case the formulas for $a'$ and $b'$ analogue to \eqref{eq:c-gen} coincide such that $a'=b'$. Moreover, we have
\[
c'=\frac{c}{2}\cdot\frac{(1-\cot^2\psi_a)\Uptheta^2-2\cot\psi_a(2a^2-c^2)}{c^2-\Uptheta^2\cot\psi_a} \,.
\]
The limiting situation $c'=0$ is obtained if and only if  $\cot^2\psi_a+\frac{2(2a^2-c^2)}{\Uptheta^2}\cot\psi_a-1=0$. 
Because in this situation $\psi_a$ obeys $180^\circ>\psi_a\geq 90^\circ$ we have $\cot\psi_a\leq 0$. Therefore, the remaining negative solution the quadratic equation is 
\[
\cot\psi_a= -\frac{\Uptheta^2}{c^2}=-\frac{\sqrt{4a^2-c}}{c}=-\frac{h_c}{\sfrac{c}{2}}\,
\]
where $h_c$ is length of $\triangle(ABC)$ over its edge $c$. If we denote half the angle of $\triangle(ABC)$ at $C$ by $\phi$, then $\cot\phi=\frac{h_c}{\sfrac{c}{2}}$ such that $\psi_a=180^\circ-\phi$. As expected, we see that in the limiting case $\psi_a$ coincides with the angle between the lines extending $a$ and $h_c$. Moreover, again as expected, $\psi_c=2\phi$ and $a'=b'=a$. 
\end{itemize}
For the explicit translation of the result to a solution of Problem \ref{prob:unbalanced}, see again the summarizing Section \ref{sec:concl}.

\section{Summary: The solutions of Problems \ref{prob:init} and \ref{prob:unbalanced}}\label{sec:concl}

\noindent{\bf The solution of the initial Problem \ref{prob:init}.}

\

We consider a generator with non-symmetric phase-to-phase voltages $U_1,U_2,U_3$ as described in Figure \ref{abb:schalt-gen}. Then the line voltages $U_1', U_2', U_3'$ are given by
\begin{align*}
\displaystyle U_1'   & =\frac{1}{\sqrt{6}}\cdot
\frac{\sqrt{3}\big(U_2^2+ U_3^2-U_1^2\big)+\Uptheta^2 }{\sqrt{ U_1^2+ U_2^2+ U_3^2 +\sqrt{3}\,\Uptheta^2}}\\
\displaystyle U_2'   & =\frac{1}{\sqrt{6}}\cdot
\frac{\sqrt{3}\big(U_3^2+ U_1^2-U_2^2\big)+\Uptheta^2 }{\sqrt{ U_1^2+ U_2^2+ U_3^2 +\sqrt{3}\,\Uptheta^2}}\\
\displaystyle  U_3'  & =\frac{1}{\sqrt{6}}\cdot
\frac{\sqrt{3}\big(U_1^2+ U_2^2-U_3^2\big)+\Uptheta^2 }{\sqrt{ U_1^2+ U_2^2+ U_3^2 +\sqrt{3}\,\Uptheta^2}} 
\end{align*}

\noindent{\bf The solution of the general Problem \ref{prob:unbalanced}}

\

Given an unbalanced star circuit according to Figure \ref{fig:nonbalanced}. 
We know the non-symmetric phase-to-phase voltages $U_1, U_2, U_3$ of the load -- or the generator. Furthermore, we know the phase differences $\psi_1,\psi_2$, and $\psi_3=360^\circ-\psi_1-\psi_2$ of the load.  Then the line voltages $ U_1', U_2', U_3'$ of the load are given by
\begin{center}
\makebox[\textwidth]{%
\begin{minipage}{2\textwidth}
\centering
$\renewcommand{\arraystretch}{2.75}
\begin{array}{l}
\displaystyle U_1'
 =\frac{1}{2}\cdot\frac{\left|\big(\cot\psi_2+\cot\psi_3\big)\big(U_2^2+U_3^2-U_1^2-\Uptheta^2\cot\psi_1\big)\right|}%
{\sqrt{U_2^2(1+\cot^2\psi_2) +U_3^2(1+\cot^2\psi_3)
-(\cot\psi_2+\cot\psi_3)\big((U_2^2+U_3^2-U_1^2)\cot\psi_1+\Uptheta^2\big)}}
\\
\displaystyle U_2'
 =\frac{1}{2}\cdot\frac{\left|\big(\cot\psi_3+\cot\psi_1\big)\big(U_3^2+U_1^2-U_2^2-\Uptheta^2\cot\psi_2\big)\right|}%
{\sqrt{U_3^2(1+\cot^2\psi_3) +U_1^2(1+\cot^2\psi_1)
-(\cot\psi_3+\cot\psi_1)\big((U_3^2+U_1^2-U_2^2)\cot\psi_2+\Uptheta^2\big)}}
\\
\displaystyle U_3'
 =\frac{1}{2}\cdot\frac{\left|\big(\cot\psi_1+\cot\psi_2\big)\big(U_1^2+U_2^2-U_3^2-\Uptheta^2\cot\psi_3\big)\right|}%
{\sqrt{U_1^2(1+\cot^2\psi_1) +U_2^2(1+\cot^2\psi_2)
-(\cot\psi_1+\cot\psi_2)\big((U_1^2+U_2^2-U_3^2)\cot\psi_3+\Uptheta^2\big)}}
\end{array}
$
\end{minipage}}
\end{center}

The special case $\psi_1=\psi_2=\psi_3=120^\circ$ coincides with the solution of our initial Problem \ref{prob:init}. This situation in particular occurs when we consider the star circuit from Figure \ref{fig:nonbalanced} to be balanced.

\noindent In both sets of formulas we use the  abbreviation
\[
\Uptheta^2=\sqrt{\big(U_1+U_2+U_3\big)\big(U_2+U_3-U_1\big)\big(U_3+U_1-U_2\big)\big(U_1+U_2-U_3\big)}
\]

{\bf Acknowledgements: }\ 
We would like to thank Christoph Reineke for his support in the formulation of some technical details. Moreover, we would like to thank the anonymous referees. Their remarks helped us to emphasize the main focus of our arguments.

\end{document}